\documentclass[12pt]{report}
\usepackage[utf8]{inputenc}
\usepackage{blindtext}
\usepackage[letterpaper, total={8.5in, 11in}, margin=2in]{geometry}
\usepackage{ragged2e}
\usepackage{blindtext}
\usepackage{graphicx}
\usepackage{amsmath}
\usepackage{amssymb}
\usepackage{gensymb}
\usepackage{array}
\date{}

\begin{document}
\centering{\huge SINR Coverage Enhancement of 6G UAV-Assisted Networks Deploying IRS\\
\vspace{24pt}
\large Mobasshir Mahbub, Raed M. Shubair}

\newpage

\RaggedRight{\textbf{\Large 1.\hspace{10pt} Introduction}}\\
\vspace{18pt}
\justifying{\noindent Sixth-generation wireless network will incorporate localization or sensing [1-20], terahertz-band signal [21-33], and various transmission technologies [34-50], etc.
UAVs have attracted a lot of interest for a variety of potential applications. Several of the most intriguing application fields are UAV-enabled connectivity [51]. UAVs can serve as airborne base stations (BSs) to deliver wireless transmission services.

IRS has evolved as a disruptive innovation, intending to control the propagation situation during wireless transmissions [52]. IRS is a medium that allows the modification of impinging transmitted signals to enhance coverage to the cell edge. The IRS concept is founded on the notion of controlling the environment through the reflection of impinging received signals and the alteration of their phase shifts.

However, UAV-aided transmissions confront coverage and connection challenges, particularly in urban areas [53]. Infrastructures, trees, vehicles, etc. may still obstruct UAV communication links to users in a coverage area.

To solve these issues, IRS-assisted UAV communications [54] have recently been envisioned as a technique to avoid barriers and improve connectivity in UAV systems. An obstructed transmission link can be repealed using the IRS by constructing several LoS links, which considerably minimizes channel attenuation.

The research aims to enhance or maximize the coverage probability of a conventional UAV-assisted communication deploying IRS. The work compares the performance of an IRS-empowered UAV-assisted communication model with a conventional UAV-aided model considering 2 GHz and millimeter wave (mmWave) carriers varying the network parameters.
\vspace{18pt}

\RaggedRight{\textbf{\Large 2.\hspace{10pt} Related Literature}}\\
\vspace{18pt}
\justifying{\noindent The work included a review of prior literature relative to UAV-assisted networks in this section.

Mahmoud et al. [55] investigated the deployment of IRS in UAV-empowered wireless communications aiming to improve the coverage of the Internet of Things (IoT) services. The work measured and compared the ergodic capacity, symbol error rate (SER), and outage probability in terms of conventional UAV-aided communications and IRS-assisted UAV communications. The results show that IRS enhances the SER, ergodic capacity, and outage performance significantly. Liu et al. [56] analyzed the downlink coverage performance of IRS-UAV-empowered non-orthogonal multiple access (NOMA) communications network. The work aimed to efficiently allocate transmit power to UAVs and users to satisfy a flexible and ubiquitous NOMA transmission. Solanki et al. [57] investigated the performance of an IRS-assisted NOMA transmission system, where the transmission of the base station (BS) is assisted by an IRS-assembled UAV. The work analyzed the outage probability of the transmission system. Wei et al. [58] performed sum rate maximization deploying IRS in UAV-assisted orthogonal frequency division multiple access (OFDMA) transmission system. Results derived that the employment of an IRS notably increases the sum-rate of UAV-assisted communication systems. However, this research as well exempted coverage probability analysis. Mozaffari et al. [59] proposed and analyzed a framework for delay-aware cell association in UAV-assisted wireless networks. However, the work exempted the deployment of the IRS.
}
\vspace{18pt}

\newpage
\RaggedRight{\textbf{\Large 3.\hspace{10pt} Measurement Model}}\\
\vspace{18pt}
\justifying{\noindent In the case of a conventional UAV-assisted communication model [59], a set of UAVs are deployed alongside ground base stations (macro and micro) to enhance the network services for user devices (UD). This work considered an IRS-enhanced UAV-assisted wireless network in which the micro base station is serving the users through an IRS embedded with a UAV for an enhanced coverage.}

\vspace{12pt}
\RaggedRight{\textit{\large A.\hspace{10pt} Conventional UAV-Assisted Network}}\\
\vspace{12pt}

\justifying In this case, the downlink received power by the users from the serving UAV is measured as follows (Eq. 1) [59], [60],

\begin{equation}
\mathsf{P}^{D(Conv.)}_{r\in \mathit{j}} = \frac{\mathsf{P}^{UAV}_{t\in \mathit{j}}}{\mathsf{K}_0 \mathsf{d}_i^2 \mu}
\end{equation}

where $\mathsf{P}^{UAV}_{t\in \mathit{j}}$ is the downlink transmit power of the serving UAV $\mathit{j}$. $\mathsf{K}_0=\left(\frac{4\pi f_c \mathsf{d}_0}{c}\right)^2$. $c$ is indicating the propagation velocity of light in $ms^-1$. $f_c$ is the frequency of the transmitted signal in Hz. $\mathsf{d}_0$ is the reference free space separation distance between transmitter (UAV) and receiver (user) and $\mathsf{d}_0$=1 m. $d_i= \sqrt{(x^{UAV}-x^{UD})^2+(y^{UAV}-y^{UD})^2+(z^{UAV}-z^{UD})^2}$ denotes the separation distance between the serving UAV at 

$(x^{UAV},y^{UAV},z^{UAV})$ and the ground user devices located at $(x^{UD},y^{UD},z^{UD})$ coordinates. $\mu$ denotes the attenuation factor.

The reference works [59] and [60] considered the transmit power of UAV, $\mathsf{P}^{UAV}_{t\in \mathit{j}}$ = 0.5 and 1 W respectively, carrier frequency, $f_c$ = 2 GHz, UAV’s altitude, $z^{UAV}$ = 200 m, and $\mu$ = 3 dB.

The downlink SINR can be calculated by the following equation (Eq. 2),

\begin{equation}
\mathsf{S}^{D(Conv.)}_{r\in \mathit{j}} = \frac{\mathsf{P}^{D(Conv.)}_{r\in \mathit{j}}}{\sum \mathsf{P}_{r\in \mathit{i}}+N_0}
\end{equation}
where $\sum \mathsf{P}_{r\in \mathit{i}}$  is the total interference received by the user devices. $N_0$ = -90 dBm is the noise power.

\vspace{12pt}
\RaggedRight{\textit{\large B.\hspace{10pt} IRS-Empowered UAV-Assisted Network}}\\
\vspace{12pt}

\justifying In the case of an IRS-UAV network, the downlink signal power received by the user devices is calculated by the equation below (Eq. 3) [61],

\begin{equation}
\mathsf{P}^{D(IRS)}_{r\in \mathit{j}} = \frac{\mathsf{d}_x \mathsf{d}_y \lambda^2 \mathsf{M}^2 \mathsf{N}^2 \mathsf{G}_t \mathsf{G}_r \mathsf{G}_{Sct.} cos\theta_t cos\theta_r \mathsf{A}^2}{(\mathit{d}_1 \mathit{d}_2)^2 64\pi^3}\mathsf{P}^{BS-IRS}_{t\in \mathit{j}}
\end{equation}
where $\mathsf{P}^{BS-IRS}_{t\in \mathit{j}}$ is the base station-to-IRS (attached with UAV) transmit power. $\mathsf{d}_x$ and $\mathsf{d}_y$ = $\lambda/2$ represent IRS scattering elements length and width. The wavelength of the signal is $\lambda$. $\mathsf{M}$ and $\mathsf{N}$ denote the numbers of transmitter-receiver elements in IRS. The scattering gain of IRS is $\mathsf{G}_{Sct.} = \frac{\mathsf{d}_x \mathsf{d}_y 4\pi}{\lambda^2}$. $\mathsf{G}_t$ and $\mathsf{G}_r$ are the transmitter-receiver gains. The transmitter (micro base station-to-IRS) and receiver (IRS-to-UD) angles are $\theta_t$ and $\theta_r$. The amplitude of the reflection is denoted by $\mathsf{A}^2$.
\begin{equation*}
\mathit{d}_1 = \sqrt{(x^{BS}-x^{IRS})^2+(y^{BS}-y^{IRS})^2+(z^{BS}-z^{IRS})^2}
\end{equation*} is the micro base station-to-IRS (attached with UAV) separation where $(x^{BS},y^{BS},z^{BS})$ and $(x^{IRS},y^{IRS},z^{IRS})$ are the coordinates, respectively.
\begin{equation*}
\mathit{d}_2 = \sqrt{(x^{IRS}-x^{UD})^2+(y^{IRS}-y^{UD})^2+(z^{IRS}-z^{UD})^2}
\end{equation*} is the IRS-to-UD separation where $(x^{IRS},y^{IRS},z^{IRS})$ and 

$(x^{UD},y^{UD},z^{UD})$ are the coordinates, respectively.

The downlink SINR in the case of an IRS-assisted UAV communication network is obtained by the following equation (Eq. 4),

\begin{equation}
\mathsf{S}^{D(IRS)}_{r\in \mathit{j}} = \frac{\mathsf{P}^{D(IRS)}_{r\in \mathit{j}}}{\sum \mathsf{P}_{r\in \mathit{i}}+N_0}
\end{equation}

\vspace{12pt}
\RaggedRight{\textit{\large C.\hspace{10pt} Probability of Coverage}}\\
\vspace{12pt}

\justifying{The user devices are said to be within the coverage of a UAV if the downlink SINR exceeds the selected threshold SINR.

\textbf{Theorem:} The probability of coverage [62] is denoted by (Eq. 5),

\begin{equation}
\begin{split}
\mathcal{P}_{cov}=1- \sum_{\mathit{j}\in \mathcal{K}} \lambda_\mathit{j} \int_{R^2} exp &\left(-\left(\frac{\mathsf{S}^{Thr.}_\mathit{j}}{\mathsf{P}_{t\in \mathit{j}}} \right)^\frac{2}{\alpha} \| \mathit{q}_\mathit{j}\|^2 \right.\sum_{\mathit{i}=1}^K \lambda_\mathit{i}\\ &\left. \mathsf{P}_{t\in \mathit{i}}^\frac{2}{\alpha}\right) \times exp \left(- \frac{\mathsf{S}^{Thr.}_\mathit{j}}{\mathsf{P}_{t\in \mathit{j}}} \| \mathit{q}_\mathit{j}\|^2 \right) d\mathit{q}_\mathit{j}
\end{split}
\end{equation}
where $\lambda_\mathit{j}$ denotes the density of the UAVs. $\mathsf{S}^{Thr.}_\mathit{j}$ is the SINR threshold. $\mathsf{P}_{t\in \mathit{j}}$ is the transmit power of the serving UAV (conv. model)/micro base station (IRS-UAV model). $\lambda_\mathit{i}$ denotes the density of the interfering base stations. $\mathsf{P}_{t\in \mathit{i}}$ indicates the transmit power of the interfering base stations. $\sigma^2$ is the noise variance. $\mathit{q}_\mathit{j}$ denote the transmitter-receiver separation. $\alpha$ is the signal attenuation factor.

\textbf{Proof:} According to the definition of the probability of coverage (Eq. 6),

\begin{equation}
\mathcal{P}_{cov} = 1-\mathbb{E} \left[ \left(\bigcup_{\mathit{j}\in \mathcal{K}} \bigcup_{\mathit{q}_{\mathit{j}} \in \phi_{\mathit{j}}} SINR > \mathsf{S}^{Thr.}_\mathit{j} \right)\right]
\end{equation}
where Eq. 6 is formulated considering the union bound and exploiting the Campbell Mecke Theorem [62] it can be represented as (Eq. 7),

\begin{equation}
\mathcal{P}_{cov} = 1-\sum_{\mathit{j}\in \mathcal{K}} \lambda_\mathit{j} \int_{R^2} \mathbb{E}\left[ \mathbb{P}\left(\frac{\mathsf{P}_{t\in \mathit{j}}}{\| \mathit{q}_\mathit{j}\|^\alpha} > \mathsf{S}^{Thr.}_\mathit{j}.I_{\mathit{q}_\mathit{j}}\right)\right]d\mathit{q}_\mathit{j}
\end{equation}
where $I_{\mathit{q}_\mathit{j}}$ denotes the interference.

Solving for (Eq. 8),
\begin{equation}
\sum_{\mathit{j}\in \mathcal{K}} \lambda_\mathit{j} \int_{R^2} \mathbb{E}\left[ \mathbb{P}\left(\frac{\mathsf{P}_{t\in \mathit{j}}}{\| \mathit{q}_\mathit{j}\|^\alpha} > \mathsf{S}^{Thr.}_\mathit{j}.I_{\mathit{q}_\mathit{j}}\right)\right]d\mathit{q}_\mathit{j}
\end{equation}

Since the propagation link is following Rayleigh fading-related distribution the equation becomes (Eq. 9),

\begin{equation}
\sum_{\mathit{j}\in \mathcal{K}} \lambda_\mathit{j} \int_{R^2} \mathcal{L}_{I_{\mathit{q}_\mathit{j}}}\left(\frac{\mathsf{S}^{Thr.}_\mathit{j}}{\mathsf{P}_{t\in \mathit{j}}}\right)exp \left(\frac{\mathsf{S}^{Thr.}_\mathit{j}\sigma^2}{\mathsf{P}_{t\in \mathit{j}}\| \mathit{q}_\mathit{j}\|^{-\alpha}}\right)
\end{equation}
where $\mathcal{L}_{I_{\mathit{q}_\mathit{j}}}$ (.) represents the interference in a Laplace transformed form. Since the tiers of the considered network are self-sufficient or independent (Eq. 10),

\begin{equation*}
\mathcal{L}_{I_{\mathit{q}_\mathit{j}}}(s) = \mathbb{E}\left[exp\left(-s\frac{\mathsf{P}_{t\in \mathit{j}}}{\| \mathit{q}_\mathit{j}\|^\alpha}\right)\right]
\end{equation*}
\begin{equation}
=\prod_{\mathit{j}\in\mathcal{K}} \mathbb{E}\left[\prod_{\mathit{q}_\mathit{j}\in \phi_\mathit{j}}exp\left(-s\frac{\mathsf{P}_{t\in \mathit{j}}}{\| \mathit{q}_\mathit{j}\|^\alpha}\right)\right]
\end{equation}

Since the propagation channels are following Rayleigh distribution the formula of Eq. 10 becomes (Eq. 11),

\begin{equation}
=\prod_{\mathit{j}\in\mathcal{K}} \mathbb{E}\left[\prod_{\mathit{q}_\mathit{j}\in \phi_\mathit{j}}\mathcal{L}_{I_{\mathit{q}_\mathit{j}}}\left(\frac{\mathsf{P}_{t\in \mathit{j}}}{\| \mathit{q}_\mathit{j}\|^\alpha}\right)\right]
\end{equation}

Applying the Poisson Point Process-aware Probability Formulating Function (Eq. 12),

\begin{equation*}
=\prod_{\mathit{j}\in\mathcal{K}}exp\left(-\lambda_\mathit{j} \int_{R^2}\left(1-\mathcal{L}_{I_{\mathit{q}_\mathit{j}}}\left(\frac{\mathsf{P}_{t\in \mathit{j}}}{\| \mathit{q}_\mathit{j}\|^\alpha}\right)\right) d_{\mathit{q}_\mathit{j}}\right)
\end{equation*}
\begin{equation}
=\prod_{\mathit{j}\in\mathcal{K}}exp\left(-\lambda_\mathit{j} \int_{R^2}\left(1-\frac{1}{\left(1+s\frac{\mathsf{P}_{t\in \mathit{j}}}{\| \mathit{q}_\mathit{j}\|^\alpha}\right)}\right) d_{\mathit{q}_\mathit{j}}\right)
\end{equation}

Deploying Euler’s Beta function and deriving the Polar coordinates from the Cartesian coordinates (Eq. 13),

\begin{equation}
\mathcal{L}_{I_{\mathit{q}_\mathit{j}}}(s) = exp\left(-s_{\mathit{q}_\mathit{j}}^{\frac{2}{\alpha}}\sum_{\mathit{j}\in \mathcal{K}}\lambda_\mathit{j} \mathsf{P}_{t\in \mathit{j}}^{\frac{2}{\alpha}}\right)
\end{equation}

Using Eq. (9) and (13) the coverage probability can be written as follows (Eq. 14),

\begin{equation}
\begin{split}
\mathcal{P}_{cov}=1- \sum_{\mathit{j}\in \mathcal{K}} \lambda_\mathit{j} \int_{R^2} exp &\left(-\left(\frac{\mathsf{S}^{Thr.}_\mathit{j}}{\mathsf{P}_{t\in \mathit{j}}} \right)^\frac{2}{\alpha} \| \mathit{q}_\mathit{j}\|^2 \right.\sum_{\mathit{i}=1}^K \lambda_\mathit{i}\\ &\left. \mathsf{P}_{t\in \mathit{i}}^\frac{2}{\alpha}\right) \times exp \left(- \frac{\mathsf{S}^{Thr.}_\mathit{j}}{\mathsf{P}_{t\in \mathit{j}}} \| \mathit{q}_\mathit{j}\|^2 \right) d\mathit{q}_\mathit{j}
\end{split}
\end{equation}

\textbf{Corollary:} The probability of coverage can be simplified as (Eq. 15),

\begin{equation}
\mathcal{P}_{cov} = 1-exp\left(-\pi\mathsf{S}_{r\in \mathit{j}}^{D{\frac{2}{\alpha}}}\frac{\lambda_\mathit{j}\mathsf{S}_\mathit{j}^{{Thr.}\frac{-2}{\alpha}}}{\sum_{\mathit{i}}\lambda_{\mathit{i}}}\right)
\end{equation}
where $\mathsf{S}_{r\in \mathit{j}}^{D}$ is the downlink SINR.}

\vspace{18pt}

\RaggedRight{\textbf{\Large 4.\hspace{10pt} Numerical Results and Discussions}}\\
\vspace{18pt}
\justifying
This section contains the numerical results derived by the equations stated in the previous section utilizing MATLAB. This work considers that the micro base stations are serving/feeding the IRS (passive reflector) attached to UAV with communication facilities to enhance the overall network coverage for users, especially cell-edge users. The reference works [59] and [60] considered that the UAV serves to extend the coverage for the user performing like a coverage extender (controlled by ground base stations). Table 1 states the measurement parameters and values.

\begin{table}[htbp]
\caption{Measurement Parameters and Values}
\begin{center}
\begin{tabular}{| m{4.5cm} | m{4.5cm}|}
\hline
\textbf{\textit{Parameters}}& \textbf{\textit{Values}}\\
\hline
Macro cell area & 1000x1000m\\
\hline
Micro cell area & 200x200m\\
\hline
Macro BS power & 30 W \\
\hline
Micro BS power & 8 W \\
\hline
Micro BS power (to IRS) & 0.1, 0.2, 4 W\\
\hline
UAV transmit power (Conv. Model) & 6 W, 8 W\\
\hline
Macro BS height & 20 m\\
\hline
Micro BS height & 10 m\\
\hline
UAV altitude & 50, 100, 200 m\\
\hline
Carrier frequencies & 2, 30, 55, 80, 100 GHz\\
\hline
Tx-Rx elements & 32, 64, 128, 256\\
\hline
Tx-Rx gain (IRS) & 20 dB [63], 14 dB [64]\\
\hline
Transmit-receive angles & 45$\degree$\\
\hline
Density of IRS & $1000⁄(\pi(100)^2 )$ per $m^2$\\
\hline
Density of Micro BS & $1000⁄(\pi(100)^2)$ per $m^2$\\
\hline
Density of Macro BS & $(1000⁄(\pi(100)^2))/5$ per $m^2$\\
\hline
Amplitude of reflection & 0.9\\
\hline
Attenuation exponent & 2\\
\hline
\end{tabular}
\label{tab1}
\end{center}
\end{table}

Fig. 1 (a), (b), (c), and (d) represent the comparative coverage probability performance analysis for the conventional and IRS-assisted UAV communication model in terms of varied measurement parameters.

\begin{figure}[htbp]
\centerline{\includegraphics[height=8.5cm, width=11.5cm]{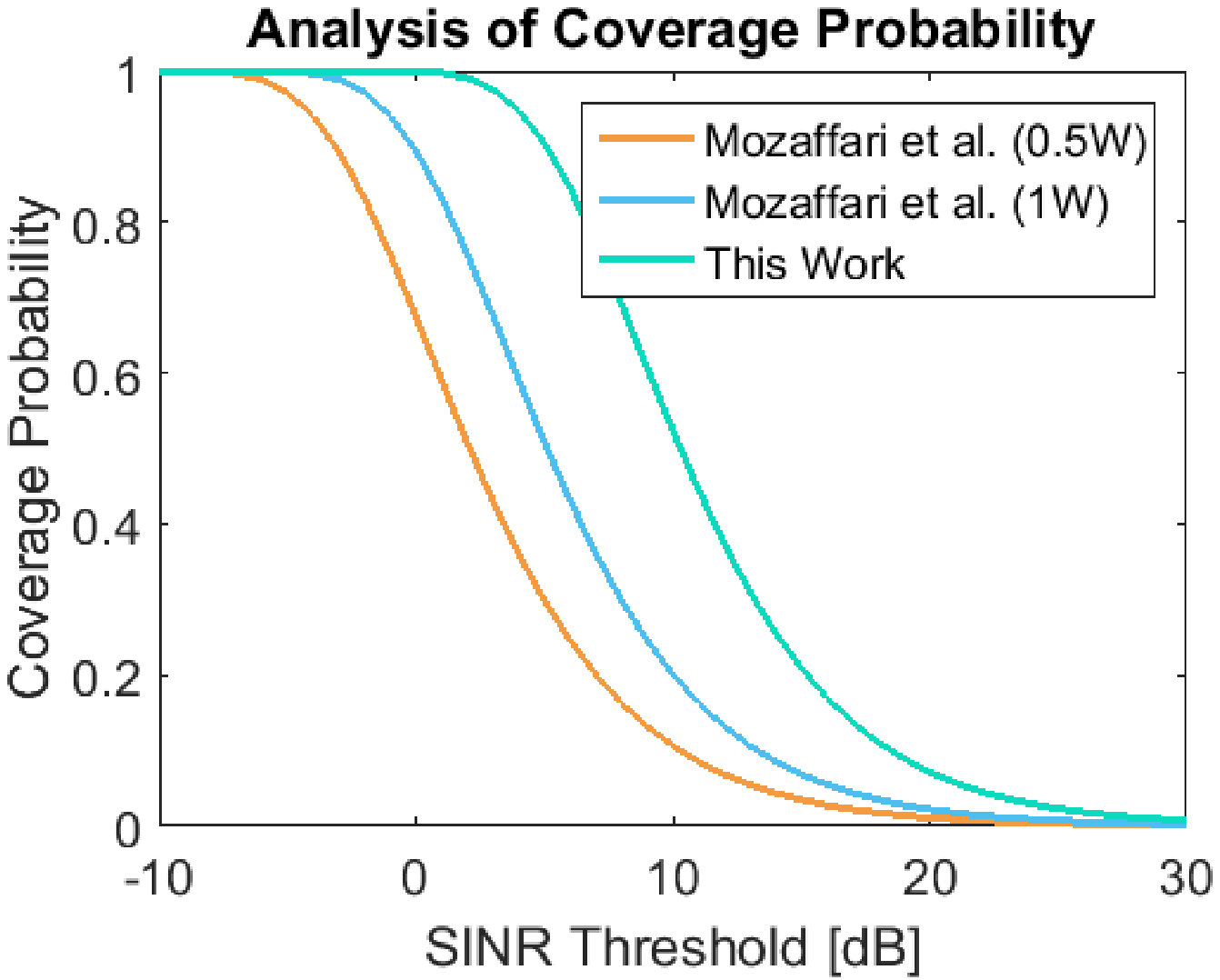}}
\vspace{3pt}
\centerline{\footnotesize{(a)}}
\label{fig}
\end{figure}

\begin{figure}[htbp]
\centerline{\includegraphics[height=8.5cm, width=11.5cm]{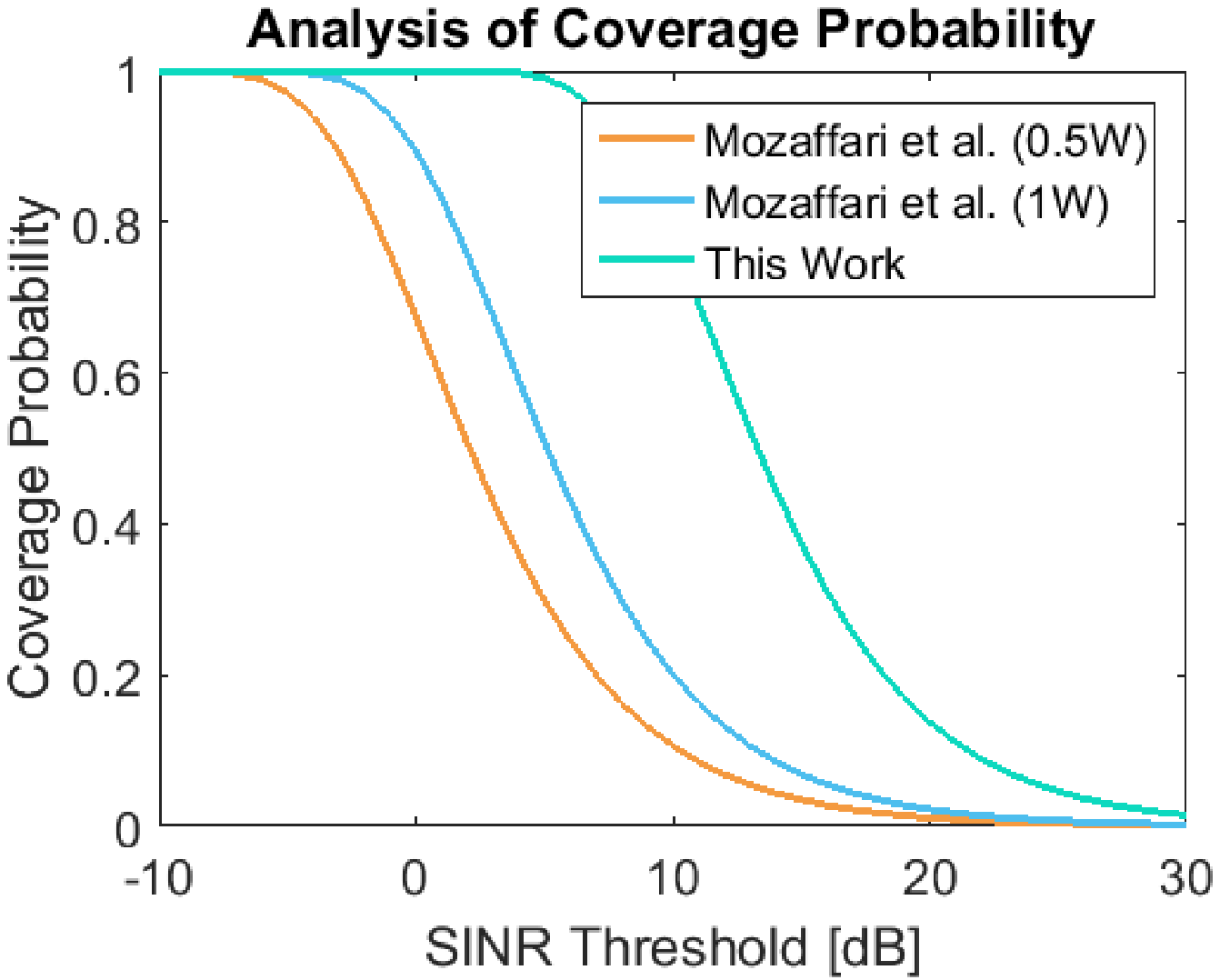}}
\vspace{3pt}
\centerline{\footnotesize{(b)}}
\label{fig}
\end{figure}

\begin{figure}[htbp]
\centerline{\includegraphics[height=8.5cm, width=11.5cm]{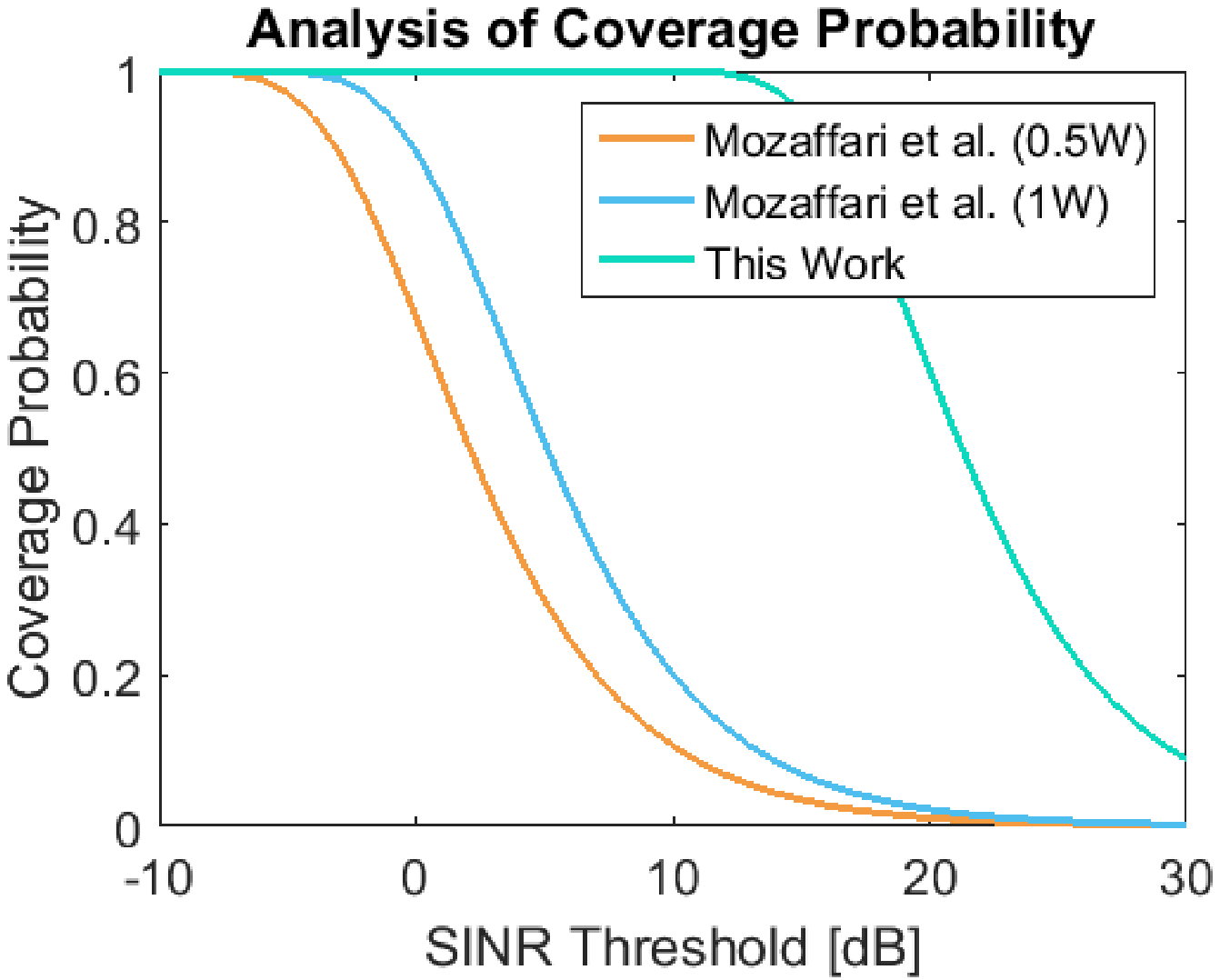}}
\vspace{3pt}
\centerline{\footnotesize{(c)}}
\label{fig}
\end{figure}

\begin{figure}[htbp]
\centerline{\includegraphics[height=8.5cm, width=11.5cm]{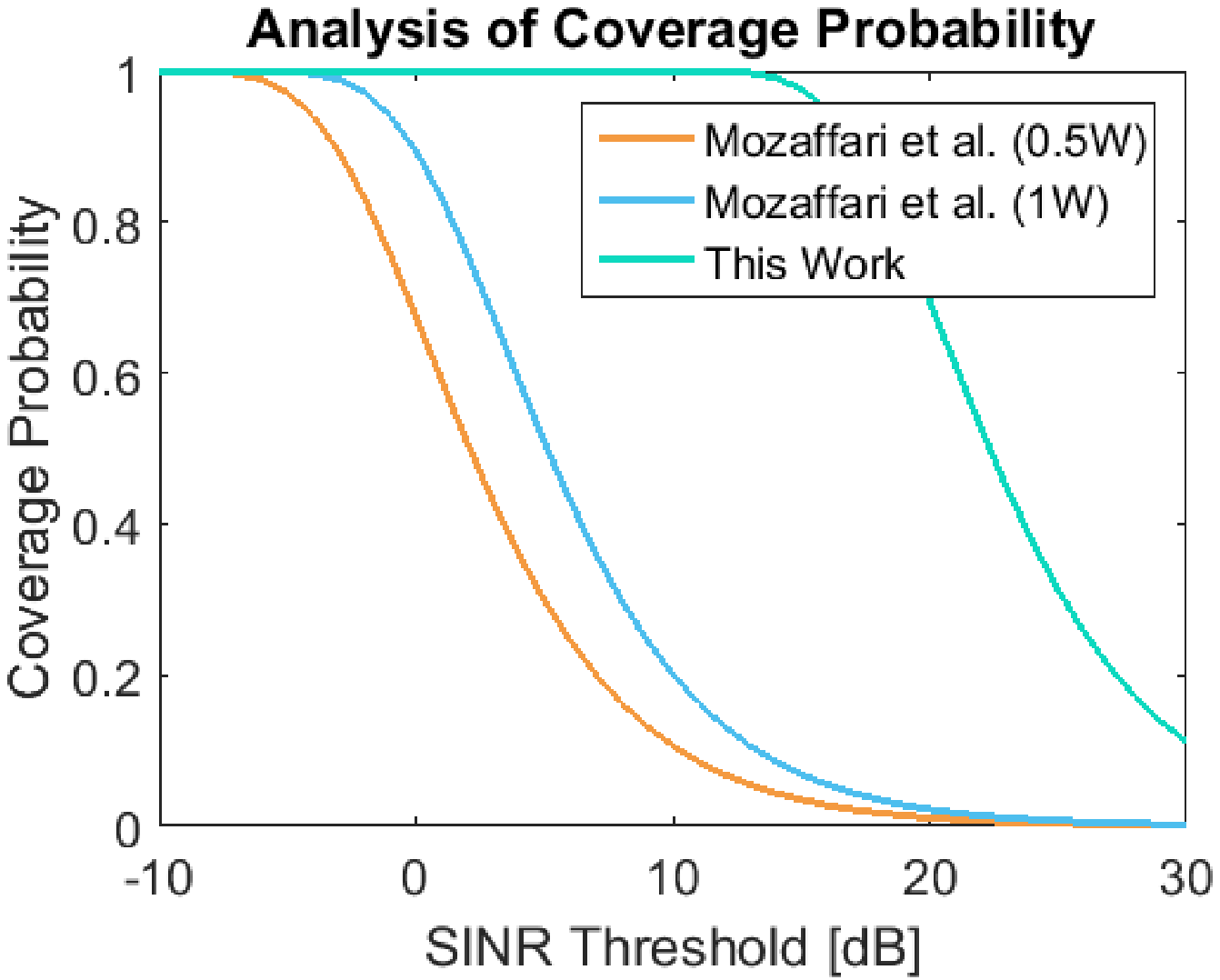}}
\vspace{3pt}
\centerline{\footnotesize{(d)}}
\caption{(a) Comparison of the coverage probabilities (transmit power 0.5 W, 1 W, 0.1 W, transmitter-receiver/IRS elements = 32, UAV altitude = 200 m), (b) Comparison of the coverage probabilities (transmit power 0.5 W, 1 W, 0.2 W, IRS elements = 32, UAV altitude = 200 m), (c) Comparison of the coverage probability (transmit power 0.5 W, 1 W, 0.2 W, IRS elements = 32, UAV altitude = 100 m), (d) Comparison of the coverage probability (transmit power 0.5 W, 1 W, 0.1 W, IRS elements = 64, UAV altitude = 200 m).}
\label{fig}
\end{figure}

Fig. 2 (a), (b), (c), and (d) illustrate the coverage probability measurements for the conventional UAV-assisted communication model for multiple mmWave carriers in terms of varied measurement parameters.

\begin{figure}[htbp]
\centerline{\includegraphics[height=8.5cm, width=11.5cm]{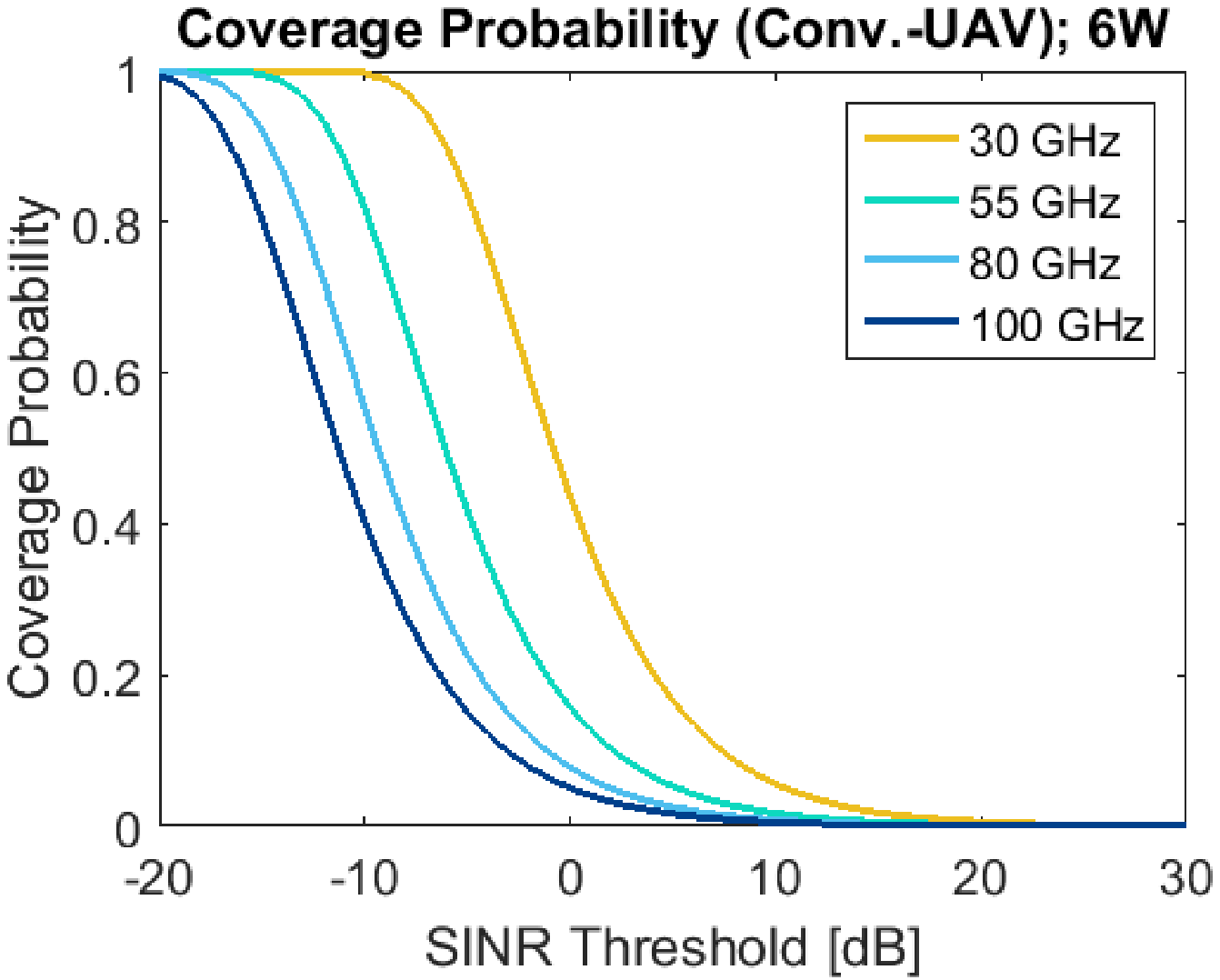}}
\vspace{3pt}
\centerline{\footnotesize{(a)}}
\label{fig}
\end{figure}

\begin{figure}[htbp]
\centerline{\includegraphics[height=8.5cm, width=11.5cm]{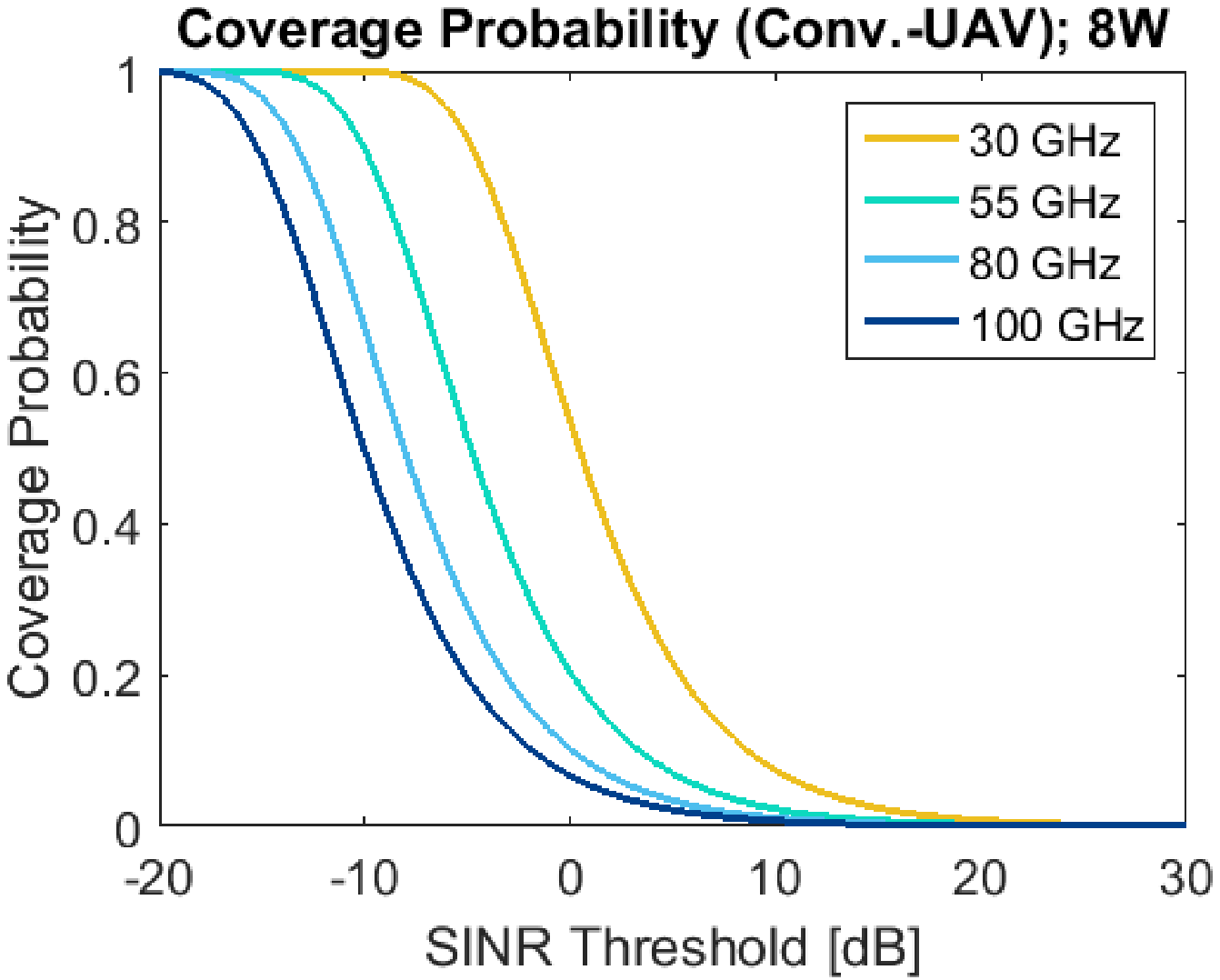}}
\vspace{3pt}
\centerline{\footnotesize{(b)}}
\label{fig}
\end{figure}

\begin{figure}[htbp]
\centerline{\includegraphics[height=8.5cm, width=11.5cm]{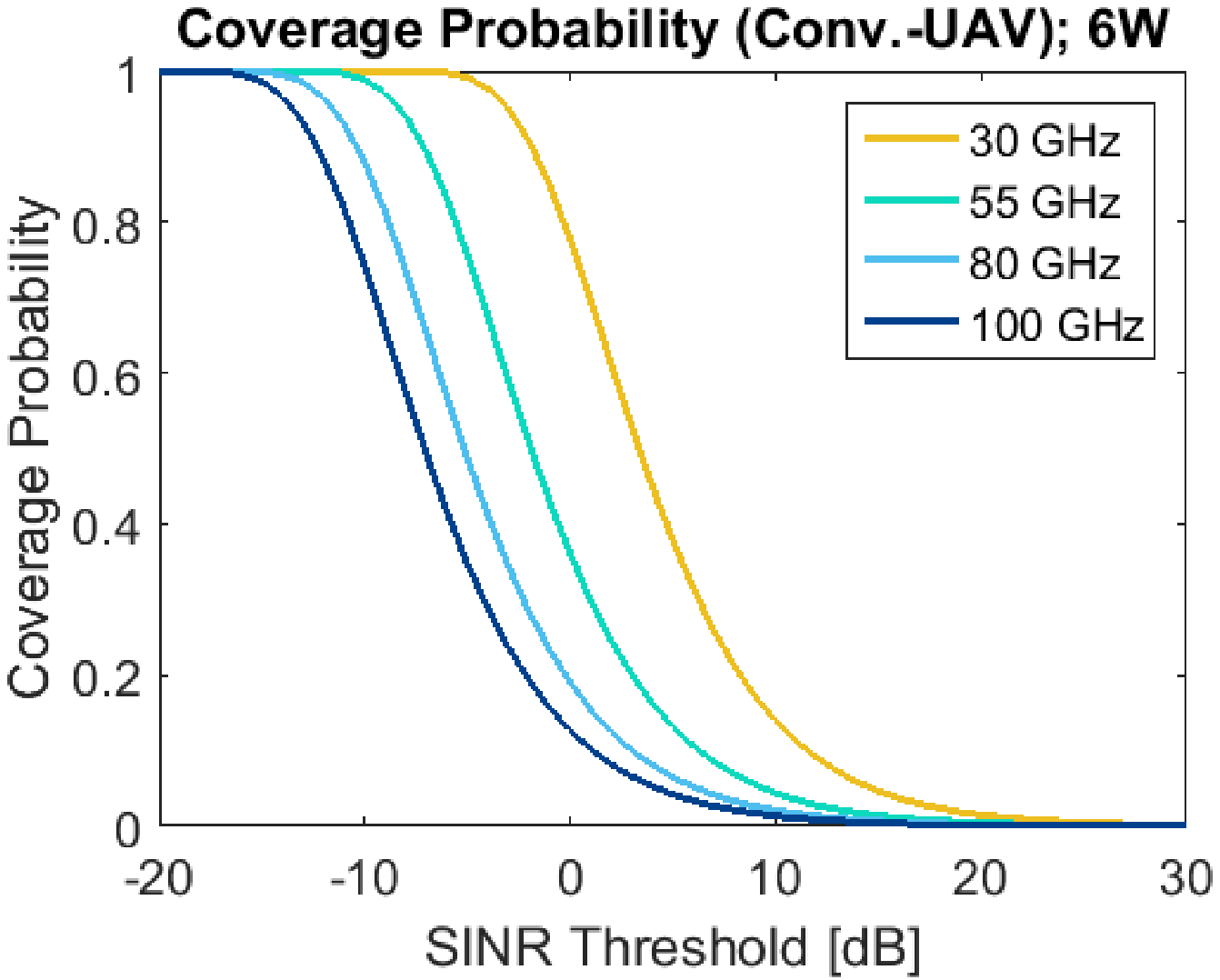}}
\vspace{3pt}
\centerline{\footnotesize{(c)}}
\label{fig}
\end{figure}

\begin{figure}[htbp]
\centerline{\includegraphics[height=8.5cm, width=11.5cm]{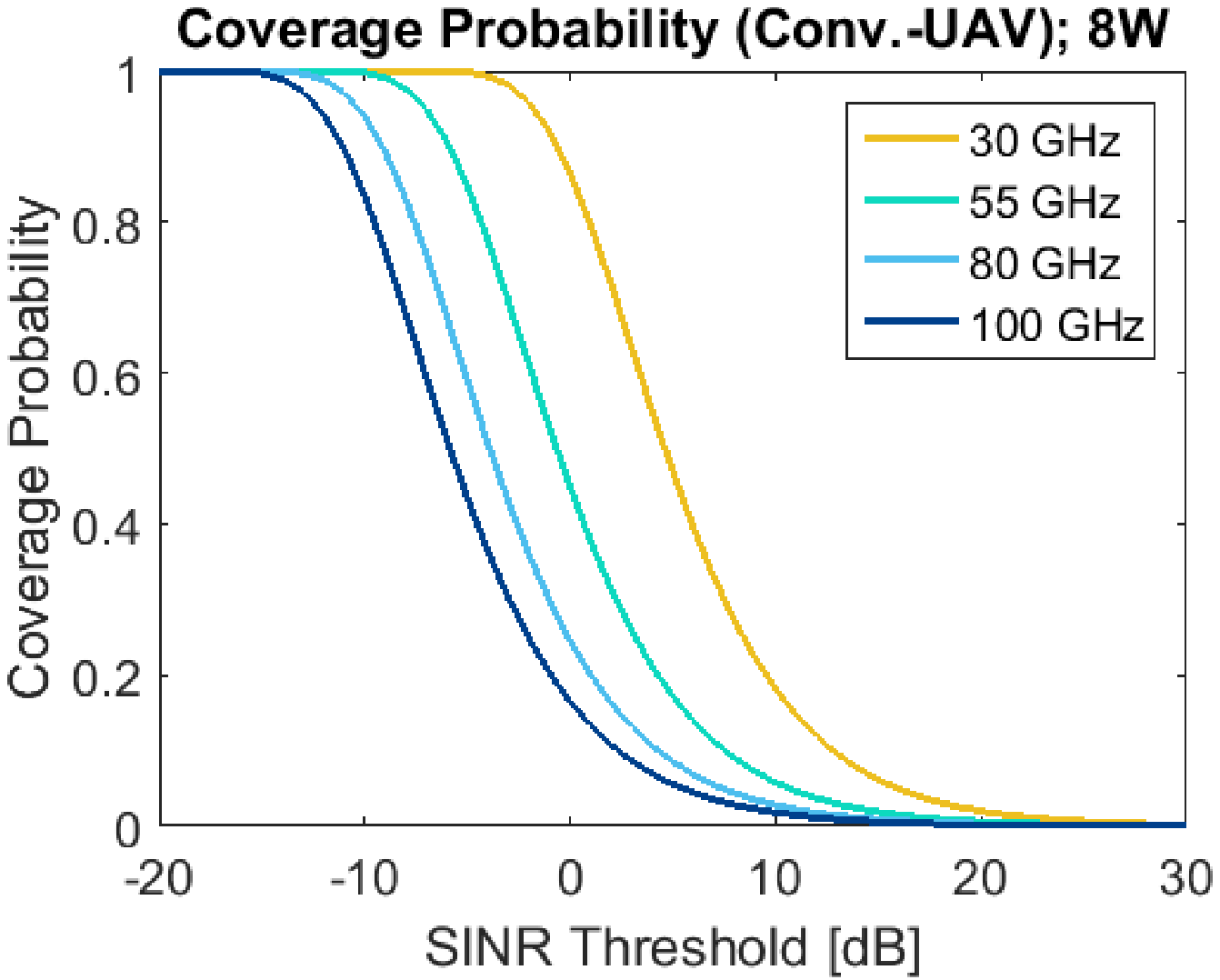}}
\vspace{3pt}
\centerline{\footnotesize{(d)}}
\caption{(a) Coverage probability for conventional UAV model (6 W, UAV altitude = 100 m), (b) Coverage probability for conventional UAV model (8 W, UAV altitude = 100 m), (c) Coverage probability for conventional UAV model (6 W, UAV altitude = 50 m), (d) Coverage probability for conventional UAV model (8 W, UAV altitude = 50 m).}
\label{fig}
\end{figure}

Fig. 3 (a), (b), (c), (d), (e), and (f) visualize the coverage probability measurements for the IRS-enhanced UAV-aided communication model for multiple mmWave carriers in terms of varied measurement parameters.

\begin{figure}[htbp]
\centerline{\includegraphics[height=8.5cm, width=11.5cm]{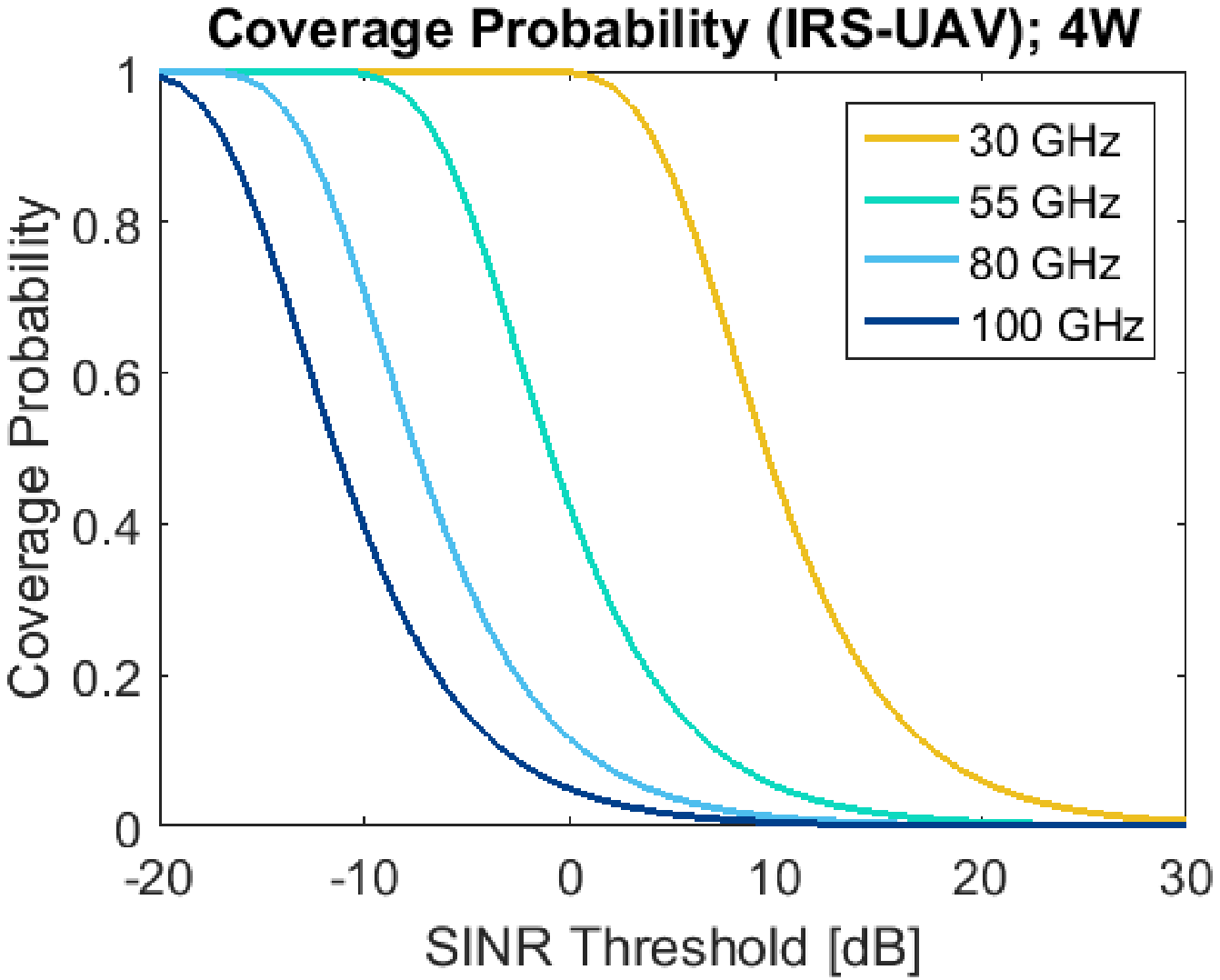}}
\vspace{3pt}
\centerline{\footnotesize{(a)}}
\label{fig}
\end{figure}

\begin{figure}[htbp]
\centerline{\includegraphics[height=8.5cm, width=11.5cm]{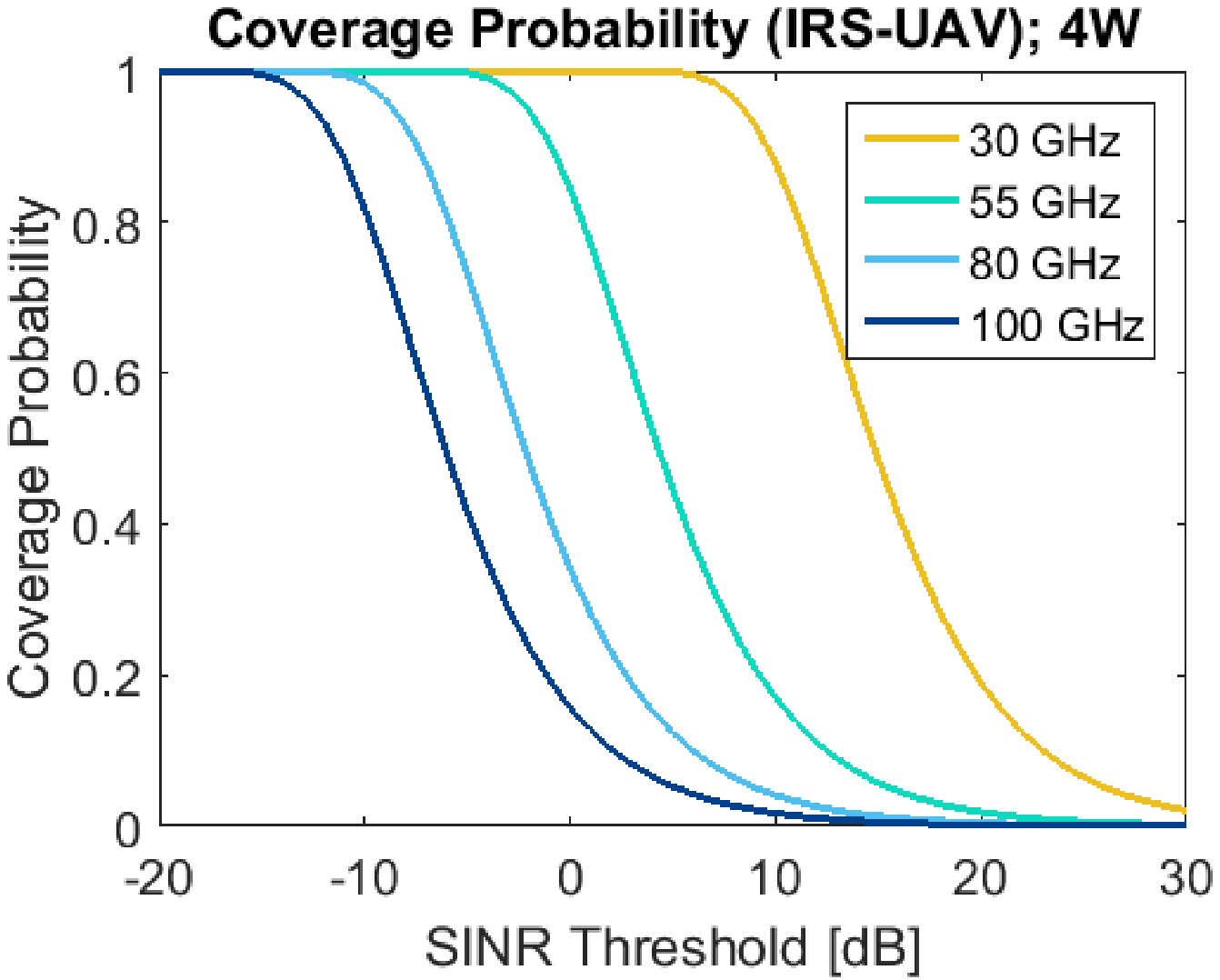}}
\vspace{3pt}
\centerline{\footnotesize{(b)}}
\label{fig}
\end{figure}

\begin{figure}[htbp]
\centerline{\includegraphics[height=8.5cm, width=11.5cm]{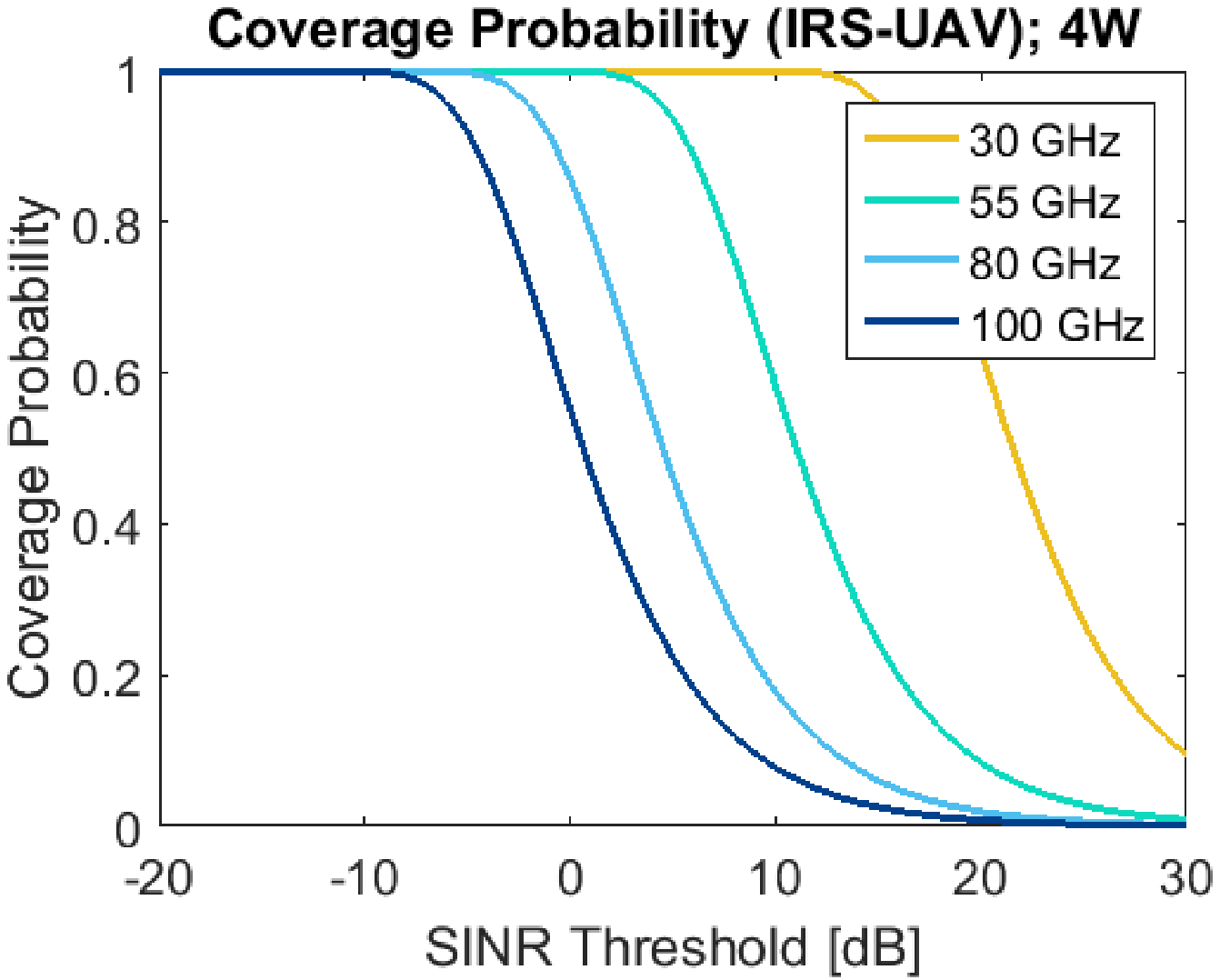}}
\vspace{3pt}
\centerline{\footnotesize{(c)}}
\label{fig}
\end{figure}

\begin{figure}[htbp]
\centerline{\includegraphics[height=8.5cm, width=11.5cm]{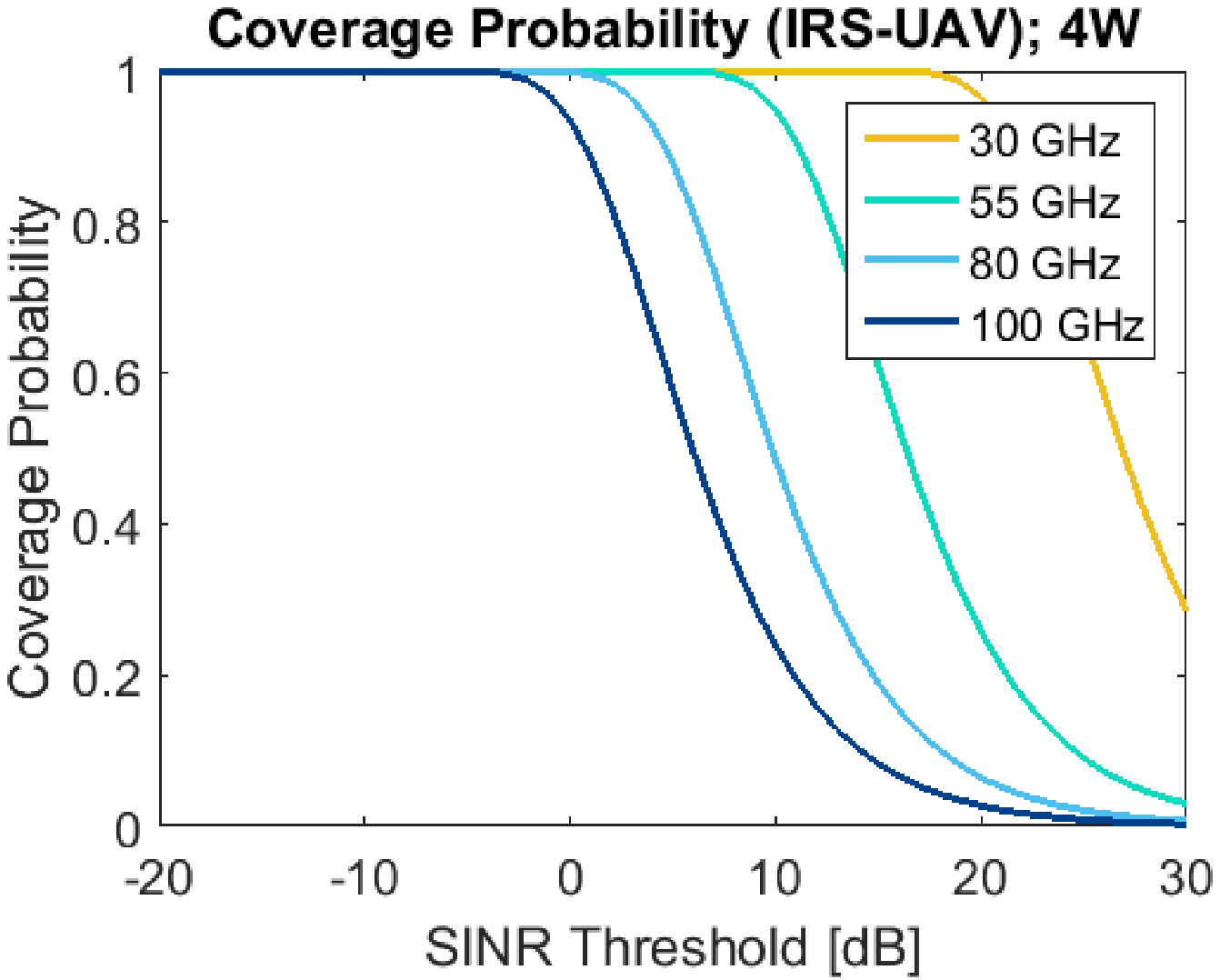}}
\vspace{3pt}
\centerline{\footnotesize{(d)}}
\caption{(a) Coverage probability for IRS-UAV model (4 W, IRS elements = 128, UAV altitude = 100 m), (b) Coverage probability for IRS-UAV model (4 W, IRS elements = 128, UAV altitude = 50 m), (c) Coverage probability for IRS-UAV model (4 W, IRS elements = 256, UAV altitude = 100 m), (d) Coverage probability for IRS-UAV model (4 W, IRS elements = 256, UAV altitude = 50 m).}
\label{fig}
\end{figure}

According to the observation of Fig. 1 (a) containing the comparative analysis among references [59] and [60] and this work, the research derived that the IRS-empowered UAV-aided communication performs better with a reduced transmit power (0.1 W) compared to the conventional UAV model. The IRS-UAV model can tolerate up to 9 dB of the SINR threshold for a coverage probability of 0.55-0.6 (denoting a median or moderately favorable coverage). On the contrary, in the case of the conventional UAV model presented in the reference works [59] and [60], respectively, 1 and 4 dB of the SINR threshold is tolerable in terms of the considered moderate or median level of coverage probability of 0.55-0.6. According to the realization of Fig. 1 (b) in the case of the IRS-UAV model if the transmit power is increased (from 0.1 W to 0.2 W) the transmission can tolerate a bit high SINR threshold i.e., up to 12 dB. From the observation of Fig. 1 (c), it is comprehensible that, if the altitude of the UAV is reduced (from 200 m to 100 m) the SINR performance improves a bit. In this case, the tolerable SINR thresholds are 6 [59], 9 [60], and 20 dB (this work). As per the realization of Fig. 1 (d), in the case of an IRS-UAV model with a transmit power of 0.1 W or 100 mW and a 200 m of UAV altitude increasing the number of elements of IRS (from 32 to 64) can enhance the tolerable SINR threshold up to 21 dB.

Since the work performed further measurements considering mmWave carriers it considered a reduced altitude of UAV with an increased power compared to the reference works [59] and [60]. Observing Fig. 2 (a) it is comprehensible that, with 6 W of transmit power and 100 m altitude of the UAV the performance of the conventional UAV is not satisfactory. In this case, the considered mmWave carriers require a very lower level of SINR threshold (-10 to 0 dB) to obtain a coverage probability of 0.55-0.6 which is not feasible in a wireless communication system. The performance is not satisfactory even with an increased transmit power of UAV, i.e., 8 W when UAV altitude is the same as previous according to Fig. 2 (b). According to the observation of Fig. 2 (c), with a reduced altitude of UAV, i.e., 50 m and 6 W of transmit power 30 GHz mmWave band can tolerate up to 2 dB of SINR threshold. However, the SINR performances of other carriers are still below the satisfactory level (-8 to -3 dB SINR threshold is required for 55, 80, and 100 GHz carriers for coverage probability of 0.55-0.6). Analyzing Fig. 2 (d) it is realizable that, with an increased transmit power of the UAV, i.e., 8 W and 50 m altitude the performance of the 30 GHz carrier improves a bit (up to 4 dB SINR threshold is tolerable). However, in this case, as well the performances of higher level mmWave carriers, e.g., 50, 80, and 100 GHz cannot be considered satisfactory (-7 to -2 dB SINR threshold is required for coverage probability of 0.55-0.6). Since the evolving 6G networks have to feature significantly higher data rates, extremely low latency, and significant reliability this kind of SINR performance of the mmWave carriers is not favorable to ensure efficient coverage to the 6G high-end user devices.

As per the interpretation of Fig. 3 (a) it is realizable that, according to the measurement parameters (4 W, IRS elements = 128, UAV altitude = 100 m) 30 GHz mmWave band exhibits better SINR performance (can tolerate up to 8 dB of SINR threshold for a coverage probability of 0.55-0.6). However, the SINR performances of other carriers are still below the satisfactory level (-12 to -2 dB SINR threshold is required for 55, 80, and 100 GHz carriers for a coverage probability of 0.55-0.6). Reduction of the UAV altitude (50 m) enhances the performance a bit namely 30 GHz can tolerate up to 14 dB of SINR threshold, 55 GHz can tolerate up to 3 dB of SINR threshold, and the performance of 80 and 100 GHz carriers still below satisfactory level according to Fig. 3 (b). With a transmit power of 4 W, 256 transmitter-receiver elements of IRS, and 100 m of altitude of UAV 30, 55, 80, and 100 GHz carriers can tolerate up to 20, 10, 3, and 0 dB of SINR thresholds, respectively, in terms of coverage probability of 0.55-0.6 by the observation of Fig. 3 (c). Analyzing Fig. 3 (d) it is comprehensible that, with a transmit power of 4 W, 256 transmitter-receiver elements of IRS, and 50 m of altitude of UAV 30, 55, 80, and 100 GHz carriers can tolerate up to 26, 15, 9, and 5 dB of SINR thresholds, respectively, in terms of coverage probability of 0.55-0.6.

\vspace{18pt}

\RaggedRight{\textbf{\Large 5.\hspace{10pt} Conclusion}}\\
\vspace{18pt}
\justifying \noindent {The research aimed to enhance the SINR coverage performance of a UAV-assisted wireless communication system deploying IRS. The work described several prior works to reflect the current research directions relative to this research issue. It formed a measurement model including the equations relative to the targeted measurement in the context of both conventional and IRS-UAV communication models. Afterward, the research analyzed and compared both of the models utilizing computer-aided measurement with MATLAB. The work derives that the deployment of IRS significantly enhances the performance of a UAV-assisted wireless network with a notable minimization of energy consumption.}
\vspace{18pt}

\RaggedRight{\textbf{\Large References}}\\
\vspace{12pt}

\justifying{
1. M. I. AlHajri, A. Goian, M. Darweesh, R. AlMemari, R. M. Shubair, L. Weruaga, and A. R. Kulaib. "Hybrid RSS-DOA technique for enhanced WSN localization in a correlated environment." In 2015 International Conference on Information and Communication Technology Research (ICTRC), pp. 238-241. IEEE, 2015.

2. M. A. Al-Nuaimi, R. M. Shubair, and K. O. Al-Midfa. "Direction of arrival estimation in wireless mobile communications using minimum variance distortionless response." In The Second International Conference on Innovations in Information Technology (IIT’05), pp. 1-5. 2005.

3. Ali Hakam, Raed M. Shubair, and Ehab Salahat. "Enhanced DOA estimation algorithms using MVDR and MUSIC." In 2013 International Conference on Current Trends in Information Technology (CTIT), pp. 172-176. IEEE, 2013.

4.	R. M. Shubair. "Robust adaptive beamforming using LMS algorithm with SMI initialization." In 2005 IEEE Antennas and Propagation Society International Symposium, vol. 4, pp. 2-5. IEEE, 2005.

5. Pradeep Kumar Singh, Bharat K. Bhargava, Marcin Paprzycki, Narottam Chand Kaushal, and Wei-Chiang Hong, eds. Handbook of wireless sensor networks: issues and challenges in current Scenario's. Vol. 1132. Berlin/Heidelberg, Germany: Springer, 2020.

6. Ebrahim M. Al-Ardi, Raed M. Shubair, and Mohammed E. Al-Mualla. "Computationally efficient DOA estimation in a multipath environment using covariance differencing and iterative spatial smoothing." In 2005 IEEE International Symposium on Circuits and Systems, pp. 3805-3808. IEEE, 2005.

7. E. M. Al-Ardi, R. M. Shubair, and M. E. Al-Mualla. "Investigation of high-resolution DOA estimation algorithms for optimal performance of smart antenna systems." (2003): 460-464.

8. M. I. AlHajri, N. Alsindi, N. T. Ali, and R. M. Shubair. "Classification of indoor environments based on spatial correlation of RF channel fingerprints." In 2016 IEEE international symposium on antennas and propagation (APSURSI), pp. 1447-1448. IEEE, 2016.

9. Goian, Mohamed I. AlHajri, Raed M. Shubair, Luis Weruaga, Ahmed Rashed Kulaib, R. AlMemari, and Muna Darweesh. "Fast detection of coherent signals using pre-conditioned root-MUSIC based on Toeplitz matrix reconstruction." In 2015 IEEE 11th International Conference on Wireless and Mobile Computing, Networking and Communications (WiMob), pp. 168-174. IEEE, 2015.

10.	R. M. Shubair, and A. Merri. "Convergence of adaptive beamforming algorithms for wireless communications." In Proc. IEEE and IFIP International Conference on Wireless and Optical Communications Networks, pp. 6-8. 2005.

11.	R. M. Shubair, A. Merri, and W. Jessmi. "Improved adaptive beamforming using a hybrid LMS/SMI approach." In Second IFIP International Conference on Wireless and Optical Communications Networks, 2005. WOCN 2005., pp. 603-606. IEEE, 2005.

12.	E. M. Al-Ardi, R. M. Shubair, and M. E. Al-Mualla. "Performance evaluation of the LMS adaptive beamforming algorithm used in smart antenna systems." In 2003 46th Midwest Symposium on Circuits and Systems, vol. 1, pp. 432-435. IEEE, 2003.

13.	Raed Shubair, and Rashid Nuaimi. "Displaced sensor array for improved signal detection under grazing incidence conditions." Progress in Electromagnetics Research 79 (2008): 427-441.

14.	M. I. AlHajri, R. M. Shubair, L. Weruaga, A. R. Kulaib, A. Goian, M. Darweesh, and R. AlMemari. "Hybrid method for enhanced detection of coherent signals using circular antenna arrays." In 2015 IEEE International Symposium on Antennas and Propagation \& USNC/URSI National Radio Science Meeting, pp. 1810-1811. IEEE, 2015.

15.	Raed M. Shubair. "Improved smart antenna design using displaced sensor array configuration." Applied Computational Electromagnetics Society Journal 22, no. 1 (2007): 83.

16.	E. M. Ardi, , R. M. Shubair, and M. E. Mualla. "Adaptive beamforming arrays for smart antenna systems: A comprehensive performance study." In IEEE Antennas and Propagation Society Symposium, 2004., vol. 3, pp. 2651-2654. IEEE, 2004.

17.	Raed M. Shubair, and Hadeel Elayan. "Enhanced WSN localization of moving nodes using a robust hybrid TDOA-PF approach." In 2015 11th International Conference on Innovations in Information Technology (IIT), pp. 122-127. IEEE, 2015.

18.	M. I. AlHajri, N. T. Ali, and R. M. Shubair. "2.4 ghz indoor channel measurements data set.” UCI Machine Learning Repository, 2018.

19.	Mohamed I. AlHajri, Nazar T. Ali, and Raed M. Shubair. "A cascaded machine learning approach for indoor classification and localization using adaptive feature selection." AI for Emerging Verticals: Human-robot computing, sensing and networking (2020): 205.

20.	Raed M. Shubair and Hadeel Elayan. "In vivo wireless body communications: State-of-the-art and future directions." In 2015 Loughborough Antennas \& Propagation Conference (LAPC), pp. 1-5. IEEE, 2015.

21.	Raed M. Shubair and Hadeel Elayan. "In vivo wireless body communications: State-of-the-art and future directions." In 2015 Loughborough Antennas \& Propagation Conference (LAPC), pp. 1-5. IEEE, 2015.

22.	Hadeel Elayan, Raed M. Shubair, Josep Miquel Jornet, and Raj Mittra. "Multi-layer intrabody terahertz wave propagation model for nanobiosensing applications." Nano communication networks 14 (2017): 9-15.

23.	Rui Zhang, Ke Yang, Akram Alomainy, Qammer H. Abbasi, Khalid Qaraqe, and Raed M. Shubair. "Modelling of the terahertz communication channel for in-vivo nano-networks in the presence of noise." In 2016 16th Mediterranean Microwave Symposium (MMS), pp. 1-4. IEEE, 2016.

24.	Maryam AlNabooda, Raed M. Shubair, Nadeen R. Rishani, and GhadahAldabbagh. "Terahertz spectroscopy and imaging for the detection and identification of illicit drugs." 2017 Sensors networks smart and emerging technologies (SENSET) (2017): 1-4.

25.	Hadeel Elayan, Cesare Stefanini, Raed M. Shubair, and Josep Miquel Jornet. "End-to-end noise model for intra-body terahertz nanoscale communication." IEEE trans. on nanobioscience 17, no. 4 (2018): 464-473.

26.	Hadeel Elayan, Raed M. Shubair, and Josep M. Jornet. "Bio-electromagnetic thz propagation modeling for in-vivo wireless nanosensor networks." In 2017 11th European Conference on Antennas and Propagation (EuCAP), pp. 426-430. IEEE, 2017.

27.	Hadeel Elayan, Raed M. Shubair, and Nawaf Almoosa. "In vivo communication in wireless body area networks." In Information Innovation Technology in Smart Cities, pp. 273-287. Springer, Singapore, 2018.

28.	Dana Bazazeh, Raed M. Shubair, and Wasim Q. Malik. "Biomarker discovery and validation for Parkinson's Disease: A machine learning approach." In 2016 International Conference on Bio-engineering for Smart Technologies (BioSMART), pp. 1-6. IEEE, 2016.

29.	S. Elmeadawy, and R. M. Shubair. "Enabling technologies for 6G future wireless communications: Opportunities and challenges. arXiv 2020." arXiv preprint arXiv:2002.06068.

30.	Hadeel Elayan, Hadeel, and Raed M. Shubair. "Towards an Intelligent Deployment of Wireless Sensor Networks." In Information Innovation Technology in Smart Cities, pp. 235-250. Springer, Singapore, 2018.

31.	Nishtha Chopra, Mike Phipott, Akram Alomainy, Qammer H. Abbasi, Khalid Qaraqe, and Raed M. Shubair. "THz time domain characterization of human skin tissue for nano-electromagnetic communication." In 2016 16th Mediterranean Microwave Symposium (MMS), pp. 1-3. IEEE, 2016.

32.	Hadeel Elayan, Raed M. Shubair, Josep M. Jornet, Asimina Kiourti, and Raj Mittra. "Graphene-Based Spiral Nanoantenna for Intrabody Communication at Terahertz." In 2018 IEEE Intl. Symposium on Antennas and Propagation \& USNC/URSI National Radio Science Meeting, pp. 799-800. IEEE, 2018.

33.	Menna El Shorbagy, Raed M. Shubair, Mohamed I. AlHajri, and Nazih Khaddaj Mallat. "On the design of millimetre-wave antennas for 5G." In 2016 16th Mediterranean Microwave Symposium (MMS), pp. 1-4. IEEE, 2016.

34.	M. Saeed Khan, A-D. Capobianco, Sajid M. Asif, Adnan Iftikhar, Benjamin D. Braaten, and Raed M. Shubair. "A pattern reconfigurable printed patch antenna." In 2016 IEEE International Symposium on Antennas and Propagation (APSURSI), pp. 2149-2150. IEEE, 2016.

35.	M. S. Khan, A. Iftikhar, A. Capobianco, R. M. Shubair, and B. Ijaz. "Pattern and frequency reconfiguration of patch antenna using PIN diodes." Microwave and Optical Technology Letters 59, no. 9 (2017): 2180-2185.

36.	M. S. Khan, A. Iftikhar, R. M. Shubair, Antonio-D. Capobianco, Benjamin D. Braaten, and Dimitris E. Anagnostou. "Eight-element compact UWB-MIMO/diversity antenna with WLAN band rejection for 3G/4G/5G communications." IEEE Open Journal of Antennas and Propagation 1 (2020): 196-206.

37.	Malak Y. ElSalamouny, and Raed M. Shubair. "Novel design of compact low-profile multi-band microstrip antennas for medical applications." In 2015 loughborough antennas \& propagation conference (LAPC), pp. 1-4. IEEE, 2015.

38.	Ala Eldin Omer, George Shaker, Safieddin Safavi-Naeini, Georges Alquié, Frédérique Deshours, Hamid Kokabi, and Raed M. Shubair. "Non-invasive real-time monitoring of glucose level using novel microwave biosensor based on triple-pole CSRR." IEEE Transactions on Biomedical Circuits and Systems 14, no. 6 (2020): 1407-1420.

39.	Saad Alharbi, Raed M. Shubair, and Asimina Kiourti. "Flexible antennas for wearable applications: Recent advances and design challenges." (2018): 484-3.

40.	R. Karli, H. Ammor, R. M. Shubair, M. I. AlHajri, and A. Hakam. "Miniature Planar Ultra-Wide-Band Microstrip Patch Antenna for Breast Cancer Detection." Skin (2016): 39.

41.	Ala Eldin Omer, George Shaker, Safieddin Safavi-Naeini, Kieu Ngo, Raed M. Shubair, Georges Alquié, F. Deshours, and Hamid Kokabi. "Multiple-cell microfluidic dielectric resonator for liquid sensing applications." IEEE Sensors Journal 21, no. 5 (2020): 6094-6104.

42.	Amjad Omar, Maram Rashad, Maryam Al-Mulla, Hussain Attia, Shaimaa Naser, Nihad Dib, and Raed M. Shubair. "Compact design of UWB CPW-fed-patch antenna using the superformula." In 2016 5th International Conference on Electronic Devices, Systems and Applications (ICEDSA), pp. 1-4. IEEE, 2016.

43.	Ahmed A. Ibrahim, and Raed M. Shubair. "Reconfigurable band-notched UWB antenna for cognitive radio applications." In 2016 16th Mediterranean Microwave Symposium (MMS), pp. 1-4. IEEE, 2016.

44.	Omar Masood Khan, Qamar Ul Islam, Raed M. Shubair, and Asimina Kiourti. "Novel multiband Flamenco fractal antenna for wearable WBAN off-body communication applications." In 2018 International Applied Computational Electromagnetics Society Symposium (ACES), pp. 1-2. IEEE, 2018.

45.	Sandip Ghosal, Arijit De, Ajay Chakrabarty, and Raed M. Shubair. "Characteristic mode analysis of slot loading in microstrip patch antenna." In 2018 IEEE International Symposium on Antennas and Propagation \& USNC/URSI National Radio Science Meeting, pp. 1523-1524. IEEE, 2018.

46.	Yazan Al-Alem, Ahmed A. Kishk, and Raed M. Shubair. "One-to-two wireless interchip communication link." IEEE Antennas and Wireless Propagation Letters 18, no. 11 (2019): 2375-2378.

47.	Nadeen R. Rishani, Raed M. Shubair, and GhadahAldabbagh. "On the design of wearable and epidermal antennas for emerging medical applications." In 2017 Sensors Networks Smart and Emerging Technologies (SENSET), pp. 1-4. IEEE, 2017.

48.	Yazan Al-Alem, Raed M. Shubair, and Ahmed Kishk. "Clock jitter correction circuit for high speed clock signals using delay units and time selection window." In 2016 16th Mediterranean Microwave Symposium (MMS), pp. 1-3. IEEE, 2016.

49.	Mikel Celaya-Echarri, Leyre Azpilicueta, Fidel Alejandro Rodríguez-Corbo, Peio Lopez-Iturri, Victoria Ramos, Mohammad Alibakhshikenari, Raed M. Shubair, and Francisco Falcone. "Towards Environmental RF-EMF Assessment of mmWave High-Node Density Complex Heterogeneous Environments." Sensors 21, no. 24 (2021): 8419.

50.	Yazan Al-Alem, Yazan, Ahmed A. Kishk, and Raed M. Shubair. "Employing EBG in Wireless Inter-chip Communication Links: Design and Performance." In 2020 IEEE International Symposium on Antennas and Propagation and North American Radio Science Meeting, pp. 1303-1304. IEEE, 2020.

51. G. Geraci et al., "What Will the Future of UAV Cellular Communications Be? A Flight from 5G to 6G," in IEEE Communications Surveys \& Tutorials, May 2022.

52. S. Gong et al., "Toward Smart Wireless Communications via Intelligent Reflecting Surfaces: A Contemporary Survey," in IEEE Communications Surveys \& Tutorials, vol. 22, no. 4, pp. 2283-2314, Fourthquarter 2020.

53. Z. Xiao et al., "A Survey on Millimeter-Wave Beamforming Enabled UAV Communications and Networking," in IEEE Communications Surveys \& Tutorials, vol. 24, no. 1, pp. 557-610, Firstquarter 2022.

54. M. Mahbub and R. M. Shubair, "Intelligent Reflecting Surfaces in UAV-Assisted 6G Networks: An Approach for Enhanced Propagation and Spectral Characteristics," 2022 IEEE International IOT, Electronics and Mechatronics Conference, 2022, pp. 1-6.

55. A. Mahmoud et al., "Intelligent Reflecting Surfaces Assisted UAV Communications for IoT Networks: Performance Analysis," in IEEE Transactions on Green Communications and Networking, vol. 5, no. 3, pp. 1029-1040, Sept. 2021.

56. C. -H. Liu, M. A. Syed and L. Wei, "Toward Ubiquitous and Flexible Coverage of UAV-IRS-Assisted NOMA Networks," 2022 IEEE Wireless Communications and Networking Conference (WCNC), 2022, pp. 1749-1754.

57. S. Solanki, J. Park and I. Lee, "On the Performance of IRS-Aided UAV Networks with NOMA," in IEEE Transactions on Vehicular Technology, April 2022.

58. Z. Wei, Y. Cai, Z. Sun, D. W. Kwan Ng and J. Yuan, "Sum-Rate Maximization for IRS-Assisted UAV OFDMA Communication Systems," GLOBECOM 2020 - 2020 IEEE Global Communications Conference, 2020, pp. 1-7.

59. M. Mozaffari, W. Saad, M. Bennis and M. Debbah, "Performance Optimization for UAV-Enabled Wireless Communications under Flight Time Constraints," GLOBECOM 2017 - 2017 IEEE Global Communications Conference, 2017, pp. 1-6.

60. M. Mozaffari, W. Saad, M. Bennis and M. Debbah, "Optimal Transport Theory for Cell Association in UAV-Enabled Cellular Networks," in IEEE Communications Letters, vol. 21, no. 9, pp. 2053-2056, Sept. 2017.

61. W. Tang et al., "Wireless Communications With Reconfigurable Intelligent Surface: Path Loss Modeling and Experimental Measurement," in IEEE Transactions on Wireless Communications, vol. 20, no. 1, pp. 421-439, Jan. 2021.

62. M. M. Fadoul, “Rate and Coverage Analysis in Multi-tier Heterogeneous Network Using Stochastic Geometry Approach,” in Ad Hoc Networks, vol.  98, March 2020.

63. M. TEKBAŞ, A. TOKTAŞ and G. ÇAKIR, "Design of a Dual Polarized mmWave Horn Antenna Using Decoupled Microstrip Line Feeder," 2020 International Conference on Electrical Engineering (ICEE), 2020, pp. 1-4.

64. F.-P. Lai, S.-Y. Mi, Y.-S. Chen, "Design and Integration of Millimeter-Wave 5G and WLAN Antennas in Perfect Full-Screen Display Smartphones" Electronics, vol. 11, no. 6, March 2022.
}

\end{document}